\documentclass[preprint,showpacs,preprintnumbers,superscriptaddress]{revtex4}
\usepackage{graphicx}
\usepackage{dcolumn}
\usepackage{bm}

\begin{document}

\title{Slow and fast micro-field components in warm and dense hydrogen
plasmas}
\author{A. Calisti}
\affiliation{PIIM, Universit\'{e} de Provence, CNRS, Centre Saint
J\'{e}r\^{o}me, 13397 Marseille Cedex 20, France.}
\author{S. Ferri}
\affiliation{PIIM, Universit\'{e} de Provence, CNRS, Centre Saint J\'{e}r\^{o}me, 13397
Marseille Cedex 20, France.}
\author{C. Moss\'{e}}
\affiliation{PIIM, Universit\'{e} de Provence, CNRS, Centre Saint J\'{e}r\^{o}me, 13397
Marseille Cedex 20, France.}
\author{B. Talin}
\affiliation{PIIM, Universit\'{e} de Provence, CNRS, Centre Saint J\'{e}r\^{o}me, 13397
Marseille Cedex 20, France.}
\author{V. Lisitsa}
\affiliation{Russian Research Center "Kurchatov Institute", Moscow, 123182, Russia.}
\author{L. Bureyeva}
\affiliation{Institute of Spectroscopy, Troitsk, Moscow region 142190, Russia.}
\author{M.A. Gigosos}
\affiliation{Departamento de \'{O}ptica y F\'{\i}sica Aplicada, Facultad de Ciencias,
Universidad de Valladolid, 47071 Valladolid, Spain.}
\author{M.A. Gonz\'{a}lez}
\affiliation{Departamento de \'{O}ptica y F\'{\i}sica Aplicada, Facultad de Ciencias,
Universidad de Valladolid, 47071 Valladolid, Spain.}
\author{T. del R\'{\i}o Gaztelurrutia}
\affiliation{Escuela Superior de Ingenieros, Ald. Urquijo s/n 48013, Bilbao, Spain.}
\author{J.W. Dufty}
\affiliation{Department of Physics University of Florida, Gainesville, FL 32611}

\begin{abstract}
The aim of this work is the investigation of the statistical
properties of local electric fields in an ion-electron two
component plasmas for coupled conditions. The stochastic fields at
a charged or at a neutral point in plasmas involve both slow and
fast fluctuation characteristics. The statistical study of these
local fields based on a direct time average is done for the first
time. For warm and
dense plasma conditions, typically $N_{e}\approx 10^{18}cm^{-3}$, $%
T_{e}\approx 1eV$, well controlled molecular dynamics (MD)
simulations of neutral hydrogen, protons and electrons have been
carried out. Relying on these \textit{ab initio} MD calculations
this work focuses on an analysis of the concepts of statistically
independent slow and fast local field components, based on the
consideration of a time averaged electric field. Large differences
are found between the results of these MD simulations and
corresponding standard results based on static screened fields.
The effects discussed are of importance for physical phenomena
connected with stochastic electric field fluctuations, e.g., for
spectral line broadening in dense plasmas.
\end{abstract}

\maketitle

The study of the local stochastic electric fields at neutral or
charged points in an homogeneous infinite plasma is of interest
for several domains. For instance they can be used as external
perturbations within semiclassical models designed to synthesize
line spectra for diagnostic purposes \cite{griem,MMM,BID,ffm}. A
more general interest comes with the study of non linear dynamics
of charges undergoing the corresponding forces. This work
addresses the simpler problem of an hydrogen plasma and the
analysis of the local field at neutral points. An equivalent study
of the local field measured at charged points is straightforward.

The mass ratio between electrons and protons results in fields
involving both fast and slow fluctuations characteristics. In
order to analyze the statistical properties of these fields, a
couple of components, a slow-fluctuation component (S) and a
fast-fluctuation component (F) is introduced,
\begin{equation}\label{eq0}
\mathbf{E}(t)=\mathbf{E}_{e}(t)+\mathbf{E}_{i}(t)=\mathbf{E}_{slow}(t)+
\mathbf{E }_{fast}(t),
\end{equation}
where $\mathbf{E}_{e}$ and $\mathbf{E}_{i}$ are the fields due to
the electrons and ions, respectively, at an arbitrary neutral
point. Our objective is to suggest a natural definition of the
slow and fast components and then determine their statistical
properties, subject to the constraints 1) their sum must be the
total field, 2) they must be statistically independent. Years ago,
S and F components have been defined in well known articles
\cite{baranger1,baranger2, hooper1,hooper2} (BMHH) and this will
be referred to as the standard model. The present work exploits
plasma molecular dynamics simulations (MD) with a more precise
definition that will be referenced as ab initio method.

The long history of plasma spectroscopy has provided indirect
interpretations of the static and dynamic properties of the local
field through their action on the relaxation of a line emitted by
the plasma \cite{alexiou,fisher}. However, the plasma and atomic
state averages required do not allow detailed  information
relevant to the splitting of the field into S and F components. In
contrast, MD simulation provides a means for direct access to such
information. The results presented here are based on the analysis
of the stochastic fluctuations of sums and differences of the
ionic and electronic local fields, $\mathbf{E}_{i}(t)$, $
\mathbf{E}_{e}(t)$ and the time averaged electron field
\begin{equation}
\overline{\mathbf{E}_{e}(t)}_{\Delta t}=\frac{1}{\Delta t}\int_{-\Delta
t/2}^{\Delta t/2}\mathbf{E}_{e}(t-t^{\prime })dt^{\prime }  \label{eq1}
\end{equation}
The mean electronic field, $\overline{\mathbf{E}_{e}(t)}_{\Delta
t}$, calculated on variable periods of time, $\Delta t$, can be
considered also as a simple measurement device, with variable
response time, averaging out on the fast fluctuation of the total
field due to the electrons. The reason for introducing this
average field is clearly to filter the total field and identify
the S and F components Accordingly, the slow component is defined
here as the slowly varying ion field plus the residual slowly
varying mean electron field
\begin{equation}
\mathbf{E}_{S,\Delta
t}(t)=\mathbf{E}_{i}(t)+\overline{\mathbf{E}_{e}(t)} _{\Delta t}.
\label{2}
\end{equation}
The fast component is the remainder $\mathbf{E}_{F,\Delta t}(t)=
\mathbf{E}(t)-\mathbf{E}_{S,\Delta
t}(t)=\mathbf{E}_{e}(t)-\overline{\mathbf{E}_{e}(t)}_{\Delta t}$.
The justification for this decomposition is provided by the
simulation results below.

MD simulation of a partially ionized hydrogen plasma gives the
local ion, electron and total fields at neutral hydrogen atoms.
The density and temperature conditions are chosen to explore
rather coupled conditions: $ N_{e}\approx 10^{18}$ cm$^{-3}$ and
$T_{e} \approx 1$~eV. Simulations of infinite systems are achieved
with periodic boundary conditions. Coulomb interactions are
shielded at a distance $\lambda\simeq s/2$, of the order of the
cubic MD cell size $s$, large enough to allow natural screening,
e.g., Debye screening, and to not affect appreciably any of the
properties investigated below. The ion-ion and electron-electron
repulsive interactions are given by a shielded Coulomb potential
$V_{12}(r)=e^{2}\mathrm{e}^{-r/\lambda}/r$. Attractive
electron-ion Coulomb interactions are regularized as follows:
$V_{ie}(r)=-e^{2}( 1-\mathrm{e}^{-r/\delta })
\mathrm{e}^{-r/\lambda}/r$, the short range regularization
parameter $\delta$ is such that the potential energy of an
electron sited on top of a proton is equal to the ionization
energy of a hydrogen atom \cite{calisti}. Regularization allowing
to implement a Coulomb attractive potential in MD simulations has
proven useful \cite{Talin,Talin2}.

Parameters of interest are the electron-electron or ion-ion
average distance, $r_{0}=(3/4\pi N_{e})^{1/3}$, defined in terms
of the electron density $N_{e}$, the mean electric field modulus
$E_{0}=e/r_{0}^{2}$, the electron thermal velocity $v_{0}
=(k_{B}T_{e}/m_{e})^{1/2}$, the electron and proton coupling
constant $\Gamma =e^{2}/r_{0}k_{B}T_{e}$ and the Debye length
$\lambda _{D}= (k_{B}T_{e}/4\pi N_{e}e^{2})^{1/2}$.

Two plasma conditions have been considered, which correspond to
$\alpha =0.8$ and $\alpha =0.4$, where $\alpha =r_{0}/\lambda
_{D}$ is the parameter introduced in BMHH. Both cases, with a same
temperature of $1$eV and coupling parameters $\Gamma \approx 0.2$
and $\Gamma \approx 0.06$, correspond to realistic experimental
plasma conditions. The number of electrons is $\approx 1000$ with
box sizes $s>3\lambda _{D}$. The same time-step $t=1.5\times
10^{-17}\mathrm{s}$ has been used for ions and electrons. This
implies that, for $\alpha =0.8$, $11600$ time-steps are required
for a proton to move across $r_{0}$ ($270$ for an electron). These
parameters give rise to intensive simulations that typically run
over a few $10^{7}$ time-steps. Times for ions and electrons to
cross the mean distance provide the ionic and electronic
characteristic times $\tau _{i}\approx 10^{-12}\mathrm{s}$ and
$\tau _{e}\approx 10^{-14}\mathrm{s}$ of fluctuations of local
fields, since the field correlation is lost, in average, once
particles have moved across the mean distance. A careful
preparation of positions and velocities of charges inside the
simulation cell guarantees a good stationarity of the total energy
all along a simulation. Typically, the relative variation of the
total energy over a simulation is smaller than $1\%$. The analysis
of the time averaged electron field introduces new complications
beyond those of current two component MD simulations and these
have been addressed with care.

\begin{figure}[tbp]
\centering
\includegraphics[width=8cm]{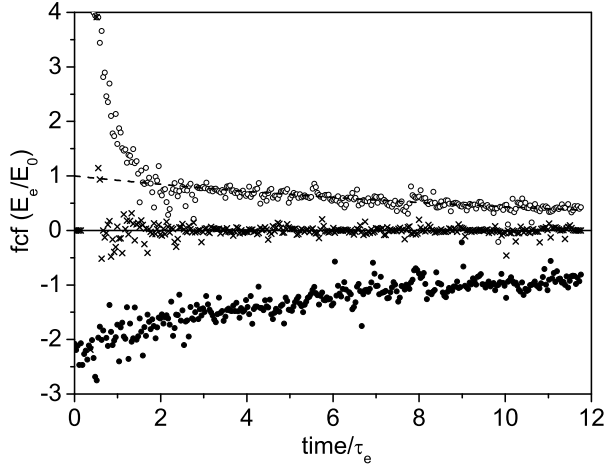}
\caption{Field correlation functions for $\protect\alpha =0.8$:
$\langle \mathbf{E}_{e}(0).\mathbf{E}_{e}(t)\rangle$, (circles);
$\langle \mathbf{E} _{F,\Delta t}(0).\mathbf{E}_{F,\Delta
t}(t)\rangle$ for $\Delta t=0.4 \protect\tau _{e}$ (crosses);
exponential fit (dash line); $\langle\mathbf{E}
_{i}(0).\mathbf{E}_{e}(t)\rangle$, (black circles).} \label{fg1}
\end{figure}

In Fig.\ref{fg1} three field correlation functions have been
plotted on the same graph in units of $\tau _{e}$, for the case
$\alpha =0.8$: $\langle \mathbf{E}_{e}(0).\mathbf{E}_{e}(t)\rangle
$ (circles), $\langle \mathbf{E} _{i}(0).\mathbf{E}_{e}(t)\rangle
$ (black circles) and $\langle \mathbf{E} _{F,\Delta
t}(0).\mathbf{E}_{F,\Delta t}(t)\rangle $ (crosses) for $\Delta
t=0.4\tau _{e}$. The symbol $\langle \rangle $ denotes a
statistical average on a relevant set of independent field
histories built, according to the present simulation method, both
at different times and for different neutral points. Clearly, in
the framework of two component plasmas, the electron field appears
inappropriate to represent the F component as the corresponding
field correlation function manifestly shows a slow de-correlation
due to electron-ion coupling mechanisms \cite{Talin2,fisher},
implying a statistical dependency between ion and electron fields.
In contrast, the correlation of $\mathbf{E}_{F,\Delta t}(t)$ with
$\Delta t=0.4\tau _{e}$ is lost over a time $<\tau _{e}$
(estimated only roughly because of the noise). The definitions for
the S and F components depend on $\Delta t$. It will be understood
hereafter that the fast and slow characteristics of fields can get
a more precise meaning by comparison to an additional time $\tau $
connected to some physical process, e.g., the relaxation of a
plasma density fluctuation or the relaxation of an atomic
radiation due to atom emitters imbedded in the plasma. The present
S and F component definitions make sense mainly for processes with
$\tau _{e}<\tau <\tau _{i}$, i.e., for components with fluctuation
times roughly connected to those of ions and electrons. For
instance, cases such that $\tau _{i}\ll \tau $, $\tau \ll \tau
_{e}$ or $\tau _{i}=\infty $ are not really considered in this
work.

The attractive interaction between ions and electrons gives rise
to an average response of the electrons to a ionic field
polarizing locally the plasma. This average response corresponds
to a natural anti-correlation mechanism i.e. $\langle
\mathbf{E}_{i}(0).\mathbf{E}_{e}(t)\rangle <0$ plotted in
Fig.\ref{fg1} and also clearly shown in \cite{fisher}. This gives
rise to the well-known model based on screening of ion-ion
interactions by the fast moving electrons. In the standard model
the slow component is taken to be a screened ion field, with a
screening length due to the electrons. A primary observation here
is that this definition of the slow component does not agree with
that of Eq.(\ref{2}). More precisely, the static screening effect
is not equivalent to the time average of the electronic motion. A
further approximation in applications of the standard model is to
replace the corresponding fast component by that due to the
electrons alone. The results below provide significant motivation
for reconsideration of those applications, particularly in light
of the early identification of screening with time averages.

In order to clarify the following comparisons it is useful to
recall a few preliminary notions. The Holtsmark function
\cite{holtsmark} gives the field distribution function (FDF)
inside an infinite space filled with a uniform random distribution
of point charges $Z=\pm 1$ at a density $N$. For instance the
Holtsmark curve is the FDF inside an infinite system of
noninteracting electrons at a density $N_{e}=N$ while the total
ion plus electron FDF, still for noninteracting charges, is given
by the Holtsmark distribution for the density $2N$. It is well
known that when the interaction between charges are switched on
the average field modulus decreases. This effect is accentuated by
increasing plasma coupling conditions.

In the two component plasma investigated here, with all the
interactions accounted for, the electronic and the ionic FDFs are
also remarkable. Due to the symmetry of charges they are the same.
In addition, when $\Delta t$ increases, the FDFs of both
$\mathbf{E}_{F,\Delta t}(t)$ and $\mathbf{E} _{S,\Delta t}(t)$
tend towards the common electronic and ionic FDF (IEFDF) as, due
to ergodicity $\lim_{\Delta
t\rightarrow\infty}\overline{\mathbf{E} _{e}(t)}_{\Delta t}=
\langle \mathbf{E}_{e}(t)\rangle = 0$. The statistical
independency of the S and F components is effective as
demonstrated in Fig.\ref{fg1}.

\begin{figure}[tbp]
\centering
\includegraphics[width=8cm]{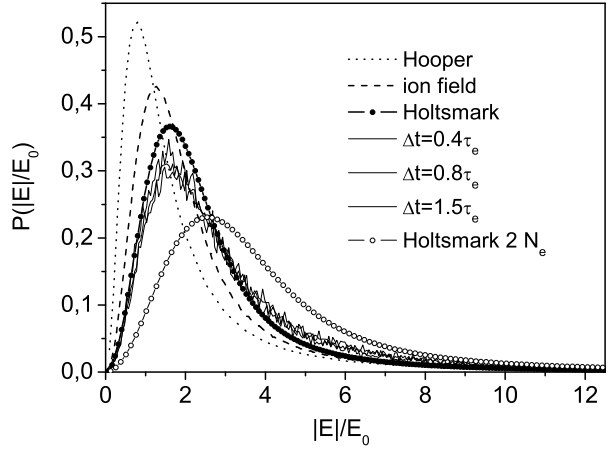}
\caption{S component field distribution functions at neutral emitters, $%
\protect\alpha=0.8$}
\label{fg2}
\end{figure}

\begin{figure}[tbp]
\centering
\includegraphics[width=8cm]{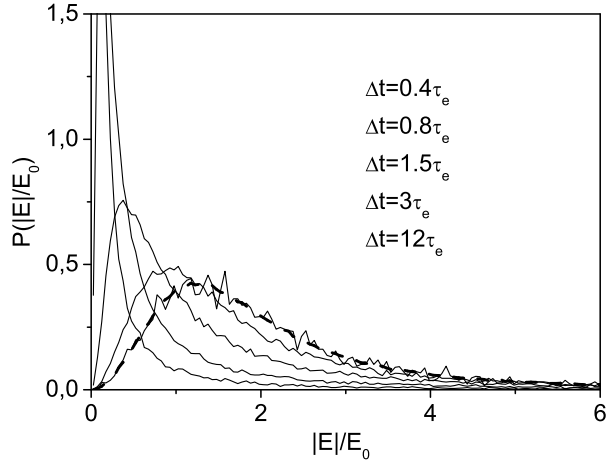}
\caption{F component field distribution functions for $\protect\alpha=0.8$, $%
\Delta t$ increases from the left to the right; dash: electron FDF}
\label{fg3}
\end{figure}

\begin{figure}[tbp]
\centering
\includegraphics[width=8cm]{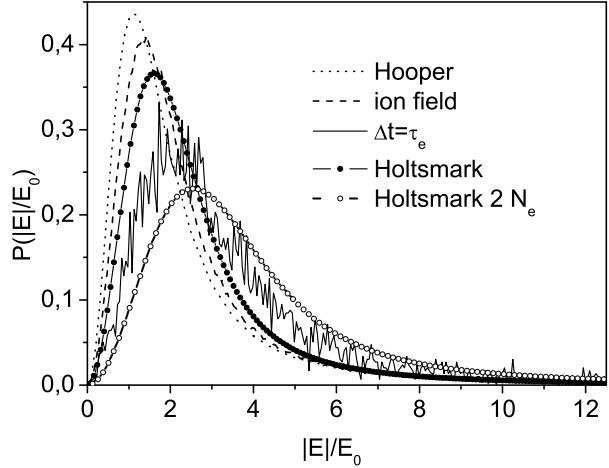}
\caption{S field distribution functions, $\protect\alpha =0.4$,
$\Delta t= \protect\tau _{e}$} \label{fg4}
\end{figure}
Figure \ref{fg2} is a typical illustration of the central problem
of this work. Three FDFs of the S component (solid lines), i.e.,
of the field $ \mathbf{E}_{S,\Delta
t}(t)=\mathbf{E}_{i}(t)+\overline{\mathbf{E}_{e}(t)} _{\Delta t}$
have been obtained with various integration times $\Delta t$. The
S component shows a small dependence on $\Delta t$ around $\tau
_{e}$, suggesting that the smallest time interval contains already
most of the contribution of electrons to the S component. The
strong field wings are located in between the two Holtsmark curves
(circles) as the slow fields result from interacting charged
particles at a density $2N_{e}$. The limit of the S component for
increasing $\Delta t$ is the ion field distribution function
(dash). As commented above, this ionic FDF resulting from
interacting ions at a density $N_{e}$ is shifted towards smaller
field modulus with respect to the Holtsmark curve (black circles).
Finally, for comparison, the field distribution function labelled
"Hooper" (dotted curve) which represent the standard model,
obtained with ionic field at emitters screened at the Debye
length, has been plotted. The large difference observed with the
present model gauges the consequences of the definition that funds
the standard model.

In the same way, the F field distribution function can be
extracted from MD simulations. In Fig.\ref{fg3} when $\Delta t$
increases the FDF is shifted from small fields to the electron
FDF, i.e., the limit obtained when $\Delta t$ increases. According
to the standard model, the electron FDF gives the F component.

An equivalent behavior is found for the case $\alpha=0.4$. Figure
\ref{fg4} shows the S field distribution function obtained with a
time-average length $ \Delta t=\tau_{e}$ together with the
Holtsmark distributions and the ion field distribution. As in the
$\alpha=0.8$ case the strong field wing of the S field
distribution function lays in between the two Holtsmark curves.
These calculations for two distinct coupling conditions show that
as expected, the Hooper S component, the ionic FDF and the
Hotsmark FDF get closer as $\Gamma$ decreases.

Coming back to the scope of this work a few points should be kept
in mind. This preliminary work exploits intensive relevant MD
sampling of stochastic local-fields at neutral points in an
hydrogen plasma for two distinct density-temperature conditions.
Studies for other conditions would be useful. However, our
objective here, i.e., the study of the statistical properties of
these sampled field in order to be able to discuss existing
models, has been reached. Due to the masses involved, local fields
undergo an intricate superposition of fast and slow fluctuations.
This suggests a rational method based on the mean electron field
for the splitting of the total field into a couple of
statistically independent S and F components whose statistical
properties are investigated separately. The procedure leading to
the S/F splitting is straightforward as it basically averages out
the fast fluctuation on variable periods of time. In contrast, the
standard BMHH procedure uses an S component rendered with an
average ionic potential involving a screening term to account for
ions dressed by electrons. The S component standard model
therefore is a purely ionic field, while that for the ab initio
method entangles both ions and those of the electrons contributing
to slow fields. Clearly, the MD ab initio approach is able to show
the anti-correlation mechanisms understood in the standard model
by the screening term implemented in the average ionic potential
but does not lead to the standard model itself, even as a limit
process. The consequence is that both approaches result in
distinct models for the FDFs.

Simple comparisons can be performed using the common IEFDF plotted
in Figs.\ref{fg2} and \ref{fg3} with dash lines. This curve i.e.,
the limit reached by $\mathbf{E}_{S,\Delta t}(t)$ and
$\mathbf{E}_{F,\Delta t}(t)$ when $\Delta t$ increases, represents
a S and a F component which are not statistically independent as
inferred from Fig. \ref{fg1}. For any $\Delta t>0$, the S
component field distributions are shifted towards stronger fields
with regard to the IEFDF. In contrast, the F component FDFs are
shifted towards smaller fields. The same occurs for the standard
model whose most probable slow field is decreased due to the Debye
screening term. In summary, in average F fields are stronger than
S fields for the standard model. The opposite is found with ab
initio methods implying that the balance between the S and F
fields results drastically different for the present and the
standard models. The unexpectedly large magnitude of the observed
differences of the FDFs should be interpreted as a warning in
favor of a careful use of the S and F components concepts, for
instance, in the context of Stark broadening of lines.

\end{document}